\documentclass[
twocolumn,reprint,
superscriptaddress,
 amsmath,amssymb,
 aps,
 prb,
]{revtex4}
\usepackage{graphicx}% Include figure files
\usepackage{dcolumn}
\usepackage{bm}
\usepackage[T1]{fontenc}
\usepackage{braket}
\usepackage{soul}
\usepackage{amssymb}
\usepackage{amsmath}

\begin{document}

\preprint{APS/123-QED}

\title{Aharonov-Bohm caging in spin-orbit coupled exciton-polariton lattices}

\author{Wei Qi}
\email{qiwei@sust.edu.cn}
\affiliation{Department of physics, Shaanxi University of Science and Technology, Xi'an 710021, China}

\date{\today}

\begin{abstract}
We study the Aharonov-Bohm (AB) caging effect in rhombic exciton-polariton lattices, with the Rashba-Dresselhaus spin-orbit coupling (RDSOC) in acting a synthetic gauge field. The effective magnetic flux through each plaquette is controlled by the orientation of the RDSOC and geometry of the rhombic lattice. The results show that the interplay of lattice geometry and the RDSOC will dramatically influence the energy band structure, furthermore, determining the transportation properties of exciton-polariton condensates. Non-Hermitian effects, which arise from the polariton intrinsic loss mechanism, on the AB caging is also discussed in detail. Meanwhile, the effect of disorder on the dynamics of AB caging is investigated, and we find that the disorder will lead to the inverse Anderson localization. We propose that using the AB caging effect allows to trap and steer the propagation of polaritons in a given parameter regime. Considering the specific example of a photonic liquid crystal microcavity to achieve our theoretical predictions, the AB caging could be switched on and off by applying an external voltage.

\end{abstract}

\maketitle

\section{Introduction}

\bigskip

Investigating the properties of localization, disorder, and transport is essential for different areas of physics and modern quantum technologies. In condensed matter physics, we know that in the presence of random disorder electron transport will be destroyed, which is called Anderson localization \cite{Anderson}. Another interesting and more controllable method is using the interplay of $\pi$-flux and lattice geometry, which can yield full localization of quantum dynamics in lattice systems, a striking interference phenomenon known as Aharonov-Bohm caging \cite{Vidal}, by which the single-particle spectrum collapses into a set of perfectly flat (dispersionless) Bloch bands. Therefore, an input excitation is decomposed in flat band states, the energy is caged, and the transport is abruptly reduced into a couple of unit cells \cite{Aravena}.
Different from the dynamic localization in other systems, the AB caging effects requiring synthetic gauge fields result in destructive interference in the rhombic lattice systems through tuning of the tunnelling amplitude and phase.

Over the past decade, there has been great interesting for synthesizing artificial gauge fields in various platforms, such as ultra-cold atoms \cite{Dalibard,Eckardt,Celi} and photonic materials \cite{KFang,YLumer}, where they constitute the basis of synthetic topological matter \cite{TOzawa} and quantum simulations \cite{IBloch}. Although the use of a magnetic flux was initially thought of for electronic lattices, this phenomenon extends to other neutral systems by the use of artificial gauge fields. With the help of synthetic magnetic fluxes induced by the synthetic gauge fields, this special flat-band localization mechanism, has spurred great interest in different areas of physics \cite{DLeykam}. AB cages were first observed in networks of conducting wires \cite{CCAbilio,CNaud}, and were recently realized in ultra-cold atom systems \cite{HangLi} and photonic lattices \cite{Gabriel} both theoretically and experimentally. In further steps, the nonlinear dynamics of AB cages \cite{MDLiberto} and nonlinear symmetry breaking of AB cages \cite{Gligoric} are discussed.

Exciton-polaritons are part-light part-matter quasiparticles formed in semiconductor microcavities \cite{HDeng,TByrnes,ICarusotto}. Many novel dynamic properties have been reported in this system arising from lattice geometry and gauged field, such as: spiraling vortices in exciton-polariton condensates \cite{Xuekai}; dynamical critical exponents in polariton quantum systems \cite{PComaron}; the flatband of a one-dimensional Lieb lattice of coupled micropillar cavities \cite{VGoblot}, exciton polaritons in a two-dimensional Lieb lattice \cite{CEWhittaker}; polariton topological insulators in flat band systems \cite{ChunyanLi}; and topological phase transition in an exciton-polariton lattice \cite{MPieczarka}. Most recently, the synthetic gauge fields have been achieved in liquid crystal exciton-polariton systems \cite{Rechnka,Gaotingge}. It is suggested that Rashba-Dresselhaus spin-orbit coupling (RDSOC) in lattices acts as a synthetic gauge field, which can be used to control the phases and magnitudes of these coupling coefficients in the lattice system \cite{Pavel}. The results present in this paper are certainly a step forward on the study of discrete dynamics \cite{Lederer,Flach}, offering a new tool for the mobility of localized wave-packets in polariton lattices.

In this article, we show that with the help of RDSOC, AB caging can be achieved in rhombic exciton-polariton lattices. We show how the orientation of RDSOC controls the localization or delocalization mechanism of polariton condensates in the lattice. Non-Hermiticity arising from the natural dissipative properties of polariton condensates \cite{Rahmani} allow the polariton system to exhibit a complex energy spectrum and wave-packet decay. Disorder is shown to lead to the inverse Anderson localization phenomenon, that is to say the presence of disorder will let the wave-packet disperse but not localize.

The paper is organized as follows. In Sec. \ref{sec2}, we present the physical models. In Sec. \ref{sec3}, Aharonov-Bohm caging in clean and Hermitian lattices is presented. The effects of non-Hermiticity on the caging dynamics is presented in Sec. \ref{sec4}. In Sec. \ref{sec5}, inverse Anderson localization in disordered lattices is discussed. Finally, In Sec. \ref{conclusion}, we give our main conclusions.

\section{Theoretical model}\label{sec2}

We consider a quasi-1D rhombic lattice with three coupled sublattices (denoted as A, B, and C) as schematically shown in Fig.~\ref{Figschematic}.
The RDSOC can be represented as a constant gauge potential that enters the tunneling coefficient as an effective phase $\beta$, where $\beta$ is proportion to the angle $\theta$, i.e. $\beta_{1,2}=\frac{\alpha a}{\hbar^{2}/2m} \cos \theta_{1,2}$ ~\cite{ZHLiu1,Aharonov}, $a$ is the link length. $\alpha$ is the amplitude and $\theta$ is the orientation of RDSOC, which allows us to tune the magnitude and sign of the tunneling coefficient. Due to the RDSOC, an effective flux $\phi=2(\beta_{1}+\beta_{2})$ is formed in each plaquette as displayed in Fig.~\ref{Figschematic}. Then the system is described by the following effective Hamiltonian:

\begin{figure}[!tbh]
\begin{center}
\rotatebox{0}{\resizebox *{6.8cm}{5.2cm} {\includegraphics
{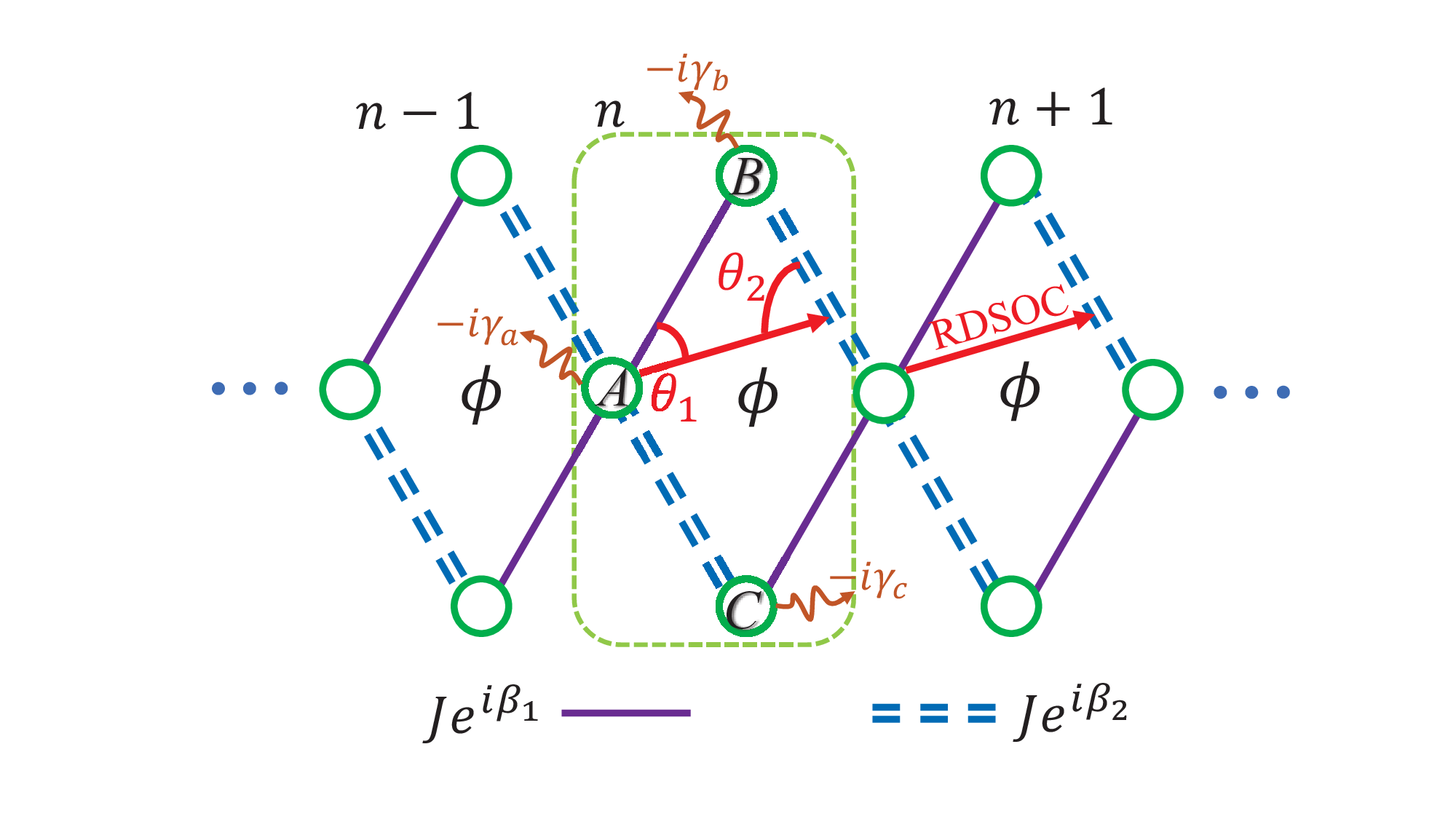}}}
\end{center}
\caption{Schematic diagram of a rhombic lattice with three sites ($A_{n}$,$B_{n}$ and $C_{n}$ per unit cell). The red arrows show the directions of spin-orbital coupling direction, and the two types of neighboring sites coupling are plotted with solid lines and double dashed lines. $\theta_{1,2}$ is represented the angle between two arms and RDSOC. A possible choice of a unit cell is marked by the green rectangle.}\label{Figschematic}
\end{figure}

\begin{eqnarray}
\begin{split}
H &= \sum_n J[(\hat{a}^{\dag}_{n}\hat{b}_{n}e^{i\beta_{1}} + \hat{a}_{n+1}\hat{b}^{\dag}_{n}e^{i\beta_{2}} \\
&+ \hat{a}^{\dag}_{n}\hat{c}_{n}e^{-i\beta_{2}} + \hat{a}_{n+1}\hat{c}^{\dag}_{n}e^{-i\beta_{1}} + H.c.)\\
&+(\Delta^{a}_{n}-i\gamma_{a})\hat{a}^{\dag}_{n}\hat{a}_{n} + (\Delta^{b}_{n}-i\gamma_{b})\hat{b}^{\dag}_{n}\hat{b}_{n} \\
&+ (\Delta^{c}_{n}-i\gamma_{c})\hat{c}^{\dag}_{n}\hat{c}_{n}],\label{1}
\end{split}
\end{eqnarray}
with $J$ the hopping constant between each site. $\hat{a}_{n}$ ($\hat{a}^{\dag}_{n}$), $\hat{b}_{n}$ ($\hat{b}^{\dag}_{n}$), $\hat{c}_{n}$ ($\hat{c}^{\dag}_{n}$) are bosonic annihilation and creation operators corresponding to the sites $A$, $B$, $C$ of the cell $n$. $\Delta^{a}_{n}$, $\Delta^{b}_{n}$, $\Delta^{c}_{n}$ and $\gamma_{a}$, $\gamma_{b}$, $\gamma_{c}$ allow for on-site disorders and polariton loss rates on each sublattice of $A$, $B$ and $C$, respectively. Under periodic boundary conditions the Hamiltonian can also be written in momentum ($k$) space:

\begin{eqnarray}
\begin{split}
H_{k} &= \sum_k J[(\hat{a}^{\dag}_{k}\hat{b}_{k}e^{i\beta_{1}} + \hat{a}_{k}\hat{b}^{\dag}_{k}e^{i\beta_{2}}e^{-ik}\\
& + \hat{a}^{\dag}_{k}\hat{c}_{k}e^{-i\beta_{2}} + \hat{a}_{k}\hat{c}^{\dag}_{k}e^{-i\beta_{1}}e^{-ik} + H.c.)]\\
& + (\Delta^{a}_{n}-i\gamma_{a})\hat{a}^{\dag}_{k}\hat{a}_{k}
 + (\Delta^{b}_{n}-i\gamma_{b})\hat{b}^{\dag}_{k}\hat{b}_{k}\\
& + (\Delta^{c}_{n}-i\gamma_{c})\hat{c}^{\dag}_{k}\hat{c}_{k},\label{2}
\end{split}
\end{eqnarray}
where $\hat{\eta}_{k}=\frac{1}{\sqrt{N}}\sum_{n}\hat{\eta}_{n}e^{ikn}$, with $\hat{\eta}_{n}=\hat{a}_{n}, \hat{b}_{n}, \hat{c}_{n}$ are bosonic annihilation and creation operators corresponding to the site A, B, C of the cell $n$, and $N$ denote the number of unit cells in the lattice.

Furthermore, from Eq. \eqref{2} we can obtain a three-band Bloch Hamiltonian in matrix form:
\begin{widetext}\label{3}
\begin{equation}
\label{Hk}
\mathcal{H}_{k} =
\begin{pmatrix}
\Delta^{a}_{n}-i\gamma_{a} & J(e^{i\beta_{1}} + e^{ik}e^{-i\beta_{2}}) & J(e^{-i\beta_{2}} +e^{ik}e^{i\beta_{1}})\\
J(e^{-i\beta_{1}}+ e^{-ik}e^{i\beta_{2}}) & \Delta^{b}_{n}-i\gamma_{b} & 0\\
J(e^{i\beta_{2}}+ e^{-ik}e^{-i\beta_{1}}) & 0 & \Delta^{c}_{n}-i\gamma_{c}
\end{pmatrix}.
\end{equation}
\end{widetext}
Solving the eigenvalues of Eq.~\eqref{Hk}, the spectrum of the system can be obtained. When the effective flux $\phi$ enclosed in each diamond plaquette is $\pi$, that is when $\beta_1+\beta_2=\pi/2$, the lattice has an entirely flat spectrum. This is known as AB caging. It is instructive to consider the dynamics of polariton condensates in such a situation. Here it is convenient to apply the mean-field approximation, by setting $A_{n}=\langle\hat{a}_{n}\rangle$, $B_{n}=\langle\hat{b}_{n}\rangle$, and $C_{n}=\langle\hat{c}_{n}\rangle$. The quantities $A_{n}$, $B_{n}$ and $C_{n}$ are the field amplitudes for the sites $A$, $B$, and $C$ in the $n$-th unit cell, $A_{n}^{*}$, $B_{n}^{*}$ and $C_{n}^{*}$ are the complex conjugate of the field amplitudes. The dynamics is then given by the Heisenberg equation of motion,
\begin{equation}
i \frac{d\eta_{n}}{dt} = \frac{\partial H}{\partial \eta_{n}^{*}}.
\end{equation}%
Then we get evolution equations of each site in the unite cell $n$,
\begin{widetext}
\begin{eqnarray}\label{4}
\left\{
\begin{aligned}
 i\dot{A}_{n} =& J(B_{n}e^{i\beta_{1}} + C_{n}e^{-i\beta_{2}} + B_{n-1}e^{-i\beta_{2}} + C_{n-1}e^{i\beta_{1}})+ (\Delta^{a}_{n}-i\gamma_{a})A_{n}, \\
 i\dot{B}_{n} =& J(A_{n}e^{-i\beta_{1}} + A_{n+1}e^{i\beta_{2}}) + (\Delta^{b}_{n}-i\gamma_{b})B_{n},\\
 i\dot{C}_{n} =& J(A_{n+1}e^{-i\beta_{1}} + A_{n}e^{i\beta_{2}}) + (\Delta^{c}_{n}-i\gamma_{c})C_{n}.
\end{aligned}
\right.
\end{eqnarray}
\end{widetext}

To further characterize localization and diffraction properties of the polariton condensate wave-packets,
it is helpful to define two characteristic quantities, namely, the inverse participation number $P^{-1}$ and wave-packet width $W$.
$P^{-1}$ is defined as:

\begin{equation}\label{5}
P^{-1}=\sum_{n}(|A_{n}|^{4}+|B_{n}|^{4}+|C_{n}|^{4}).
\end{equation}%
The inverse participation number is always smaller than or equal to 1 and it gives a measure of the
number of sites where condensates are confined. For example, if we have $P=1$ then exciton-polaritons are confined to a single site,
and if $P\sim m$ the polaritons are confined to a cluster of $m$ sites. Another useful quantity is the average square width $W^{2}$, defined as
\begin{equation}\label{6}
W^{2}=\frac{\langle x^{2}\rangle-\langle x\rangle^{2}}{N^{2}},
\end{equation}%
with $\langle x\rangle=\sum_{n}n(|A_{n}|^{2}+|B_{n}|^{2}+|C_{n}|^{2})$ and $\langle x^{2}\rangle=\sum_{n}n^{2}(|A_{n}|^{2}+|B_{n}|^{2}+|C_{n}|^{2})$.
The average width $W$ is useful to characterize how signals or wave-packets injected into the system disperse: it equals zero in the presence of caging and grows over time if dispersion is present.

\section{Aharonov-Bohm caging in clean and Hermitian lattices}\label{sec3}

\begin{figure}[!tbh]
\begin{center}
\rotatebox{0}{\resizebox *{8.8cm}{4.2cm} {\includegraphics
{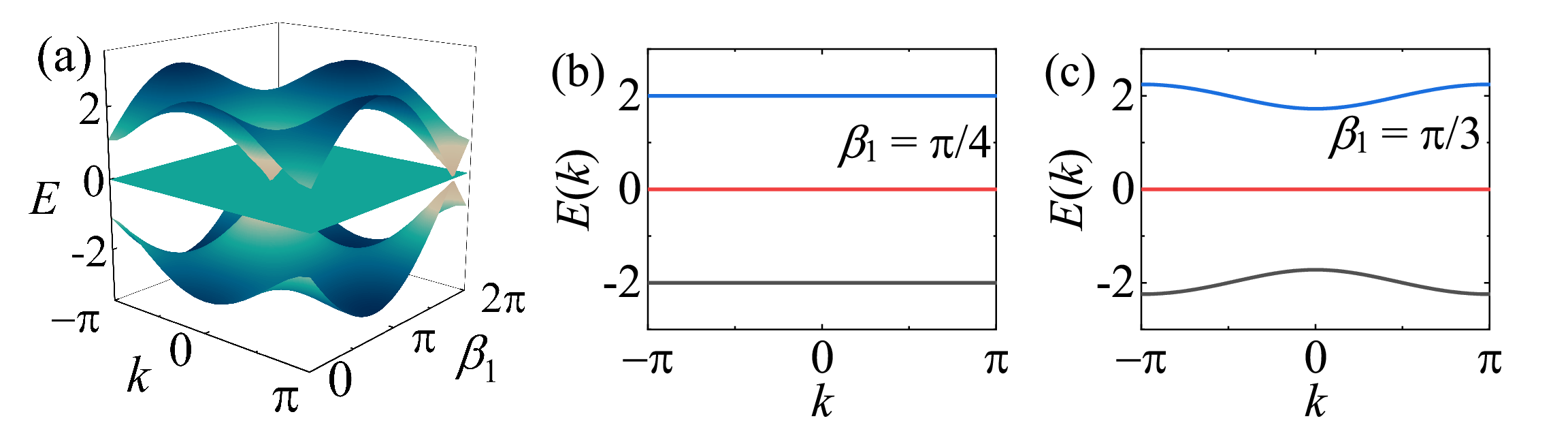}}}
\end{center}
\caption{(a) Spectrum of $\mathcal{H}(k,\beta_{1})$ in matrix of Eq.~\eqref{Hk}, (b) the flat band of $E$ with respect to $k$ with $\beta_{1}=\pi/4$ and (c) the dispersive band $E(k)$ for $\beta_{1}=\pi/3$. With $J=-1$, $\beta_{2}=\pi/4$, $\Delta^{a}_{n}=\Delta^{b}_{n}=\Delta^{c}_{n}=0$, and $\gamma_{a}=\gamma_{c}=\gamma_{b}=0$.}\label{Fig1a}
\end{figure}

First, we consider the ideal case of a clean lattice without disorder, and no losses. In this situation, the parameters in Eq. \eqref{1} will be $\Delta^{a}_{n}=\Delta^{b}_{n}=\Delta^{c}_{n}=0$ and $\gamma_{a}=\gamma_{b}=\gamma_{c}=0$. The energy spectrum is displayed in Fig.~\ref{Fig1a} (a). In Fig.~\ref{Fig1a} (b), we choose $\beta_{1}=\beta_{2}=\pi/4$ that satisfies the AB caging condition $\beta_{1}+\beta_{2}=\pi/2$. We can see in this condition, the three energy bands are all flat, representing AB caging. However, when the AB caging condition is broken, that means $\beta_{1}+\beta_{2}\neq\pi/2$, two of the flat bands will become the dispersive, as shown in Fig.~\ref{Fig1a} (c).

\begin{figure}[!tbh]
\begin{center}
\rotatebox{0}{\resizebox *{7.8cm}{4.2cm} {\includegraphics
{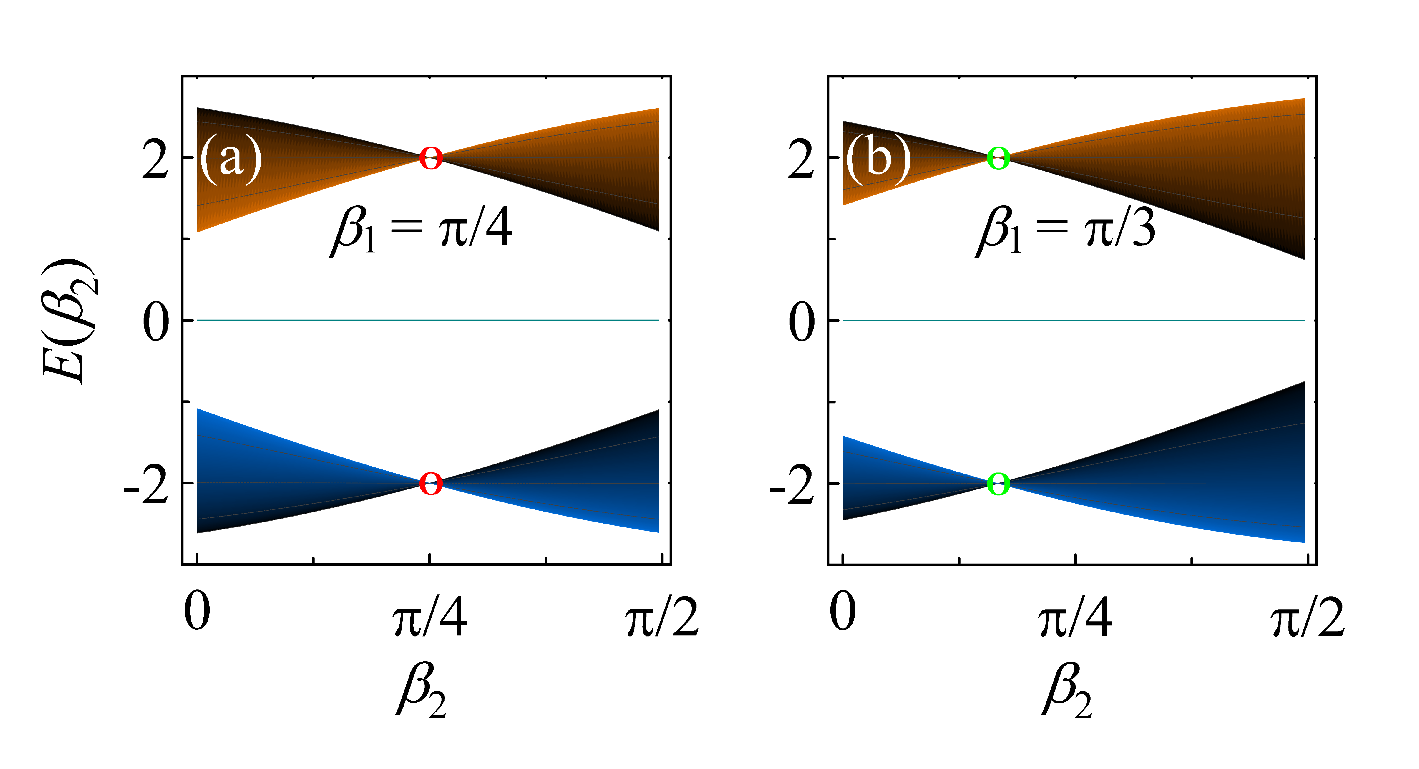}}}
\end{center}
\caption{Energy bands as a function of $\beta_{2}$ with fixed values of $\beta_{1}=\pi/4$ (a) and $\beta_1=\pi/3$ (b). The red and green circles indicate the points where the AB caging occurs. Other lattice parameters: J=-1, $\Delta^{a}_{n}=\Delta^{b}_{n}=\Delta^{c}_{n}=0$, and $\gamma_{a}=\gamma_{c}=\gamma_{b}=0$.}\label{Fig1b}
\end{figure}

It is instructive to consider how the orientation of RDSOC can influence the band structure, potentially changing also the dynamics of the system in transition to and from the AB caging situation. In Fig.~\ref{Fig1b}, the energy bands as a function of $\beta_{2}$ are shown in (a) and (b) for a fixed $\beta_{1}=\pi/4$ and $\beta_{1}=\pi/3$, respectively. It is clear that the orientation of RDSOC will dramatically change the band structure, and there exists an energy degenerate point in Fig.~\ref{Fig1b} (a) circled with red line and (b) circled with green line, which correspond to the angles $\beta_{2}=\pi/4$ (a) and $\beta_{2}=\pi/6$ (b), coinciding with the AB caging condition $\beta_{1}+\beta_{2}=\pi/2$. In the following, we mainly choose the parameters $\beta_{2}=\beta_{1}=\pi/4$, that is, the case shown in Fig.~\ref{Fig1b} (a) as an example to illustrate the caging dynamics and related properties.

\begin{figure}[!tbh]
\begin{center}
\rotatebox{0}{\resizebox *{8.8cm}{6.8cm} {\includegraphics
{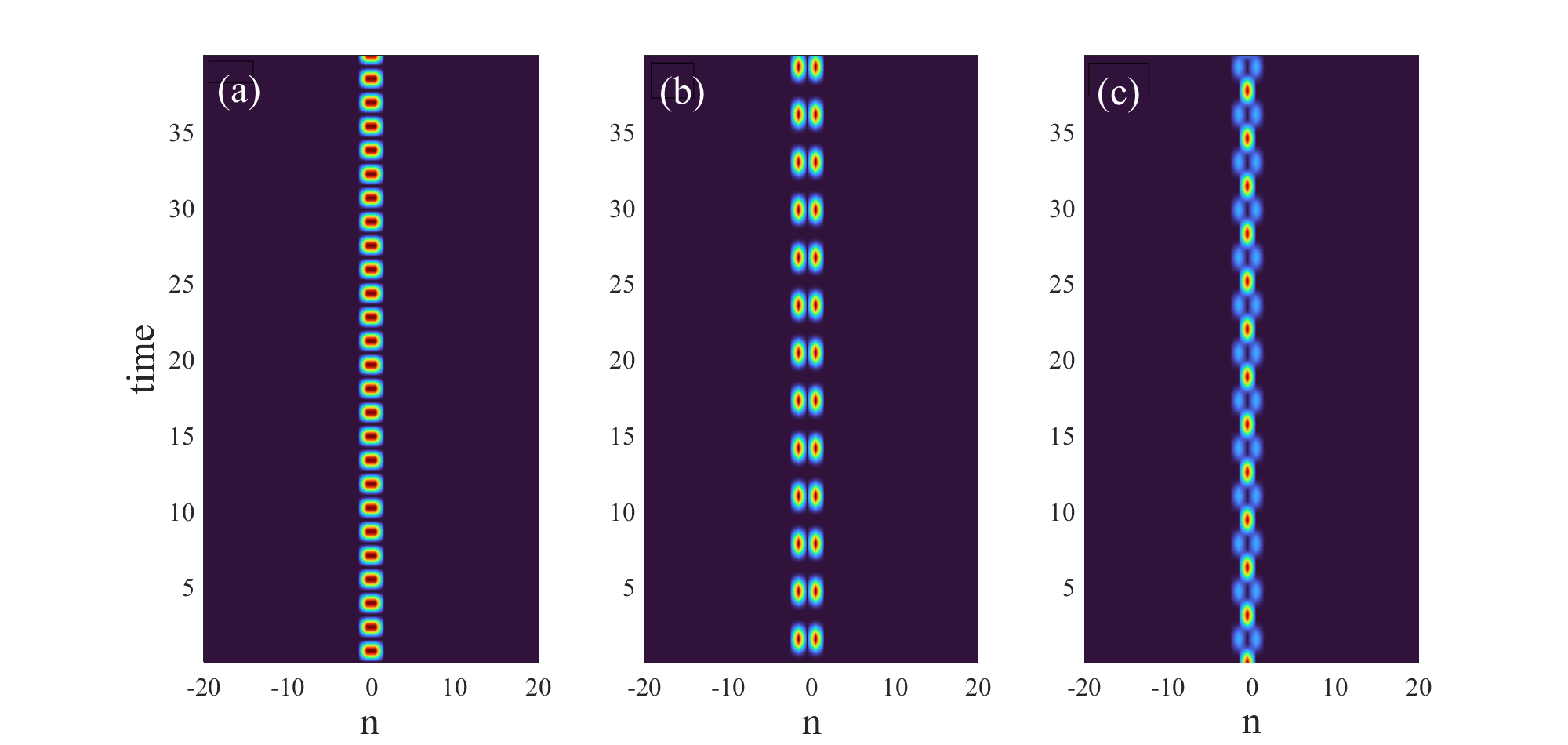}}}
\end{center}
\caption{AB caging dynamics of each sites ($A_{n}$,$B_{n}$ and $C_{n}$ in per unit cell) as show in (a), (b) and (c), respectively. With $\beta_{1}$=$\beta_{2}$=$\pi/4$, other lattice parameters: J=-1, $\Delta^{a}_{n}=\Delta^{b}_{n}=\Delta^{c}_{n}=0$, and $\gamma_{a}=\gamma_{c}=\gamma_{b}=0$.}\label{Fig1c}
\end{figure}

\begin{figure}[!tbh]
\begin{center}
\rotatebox{0}{\resizebox *{8.8cm}{5.2cm} {\includegraphics
{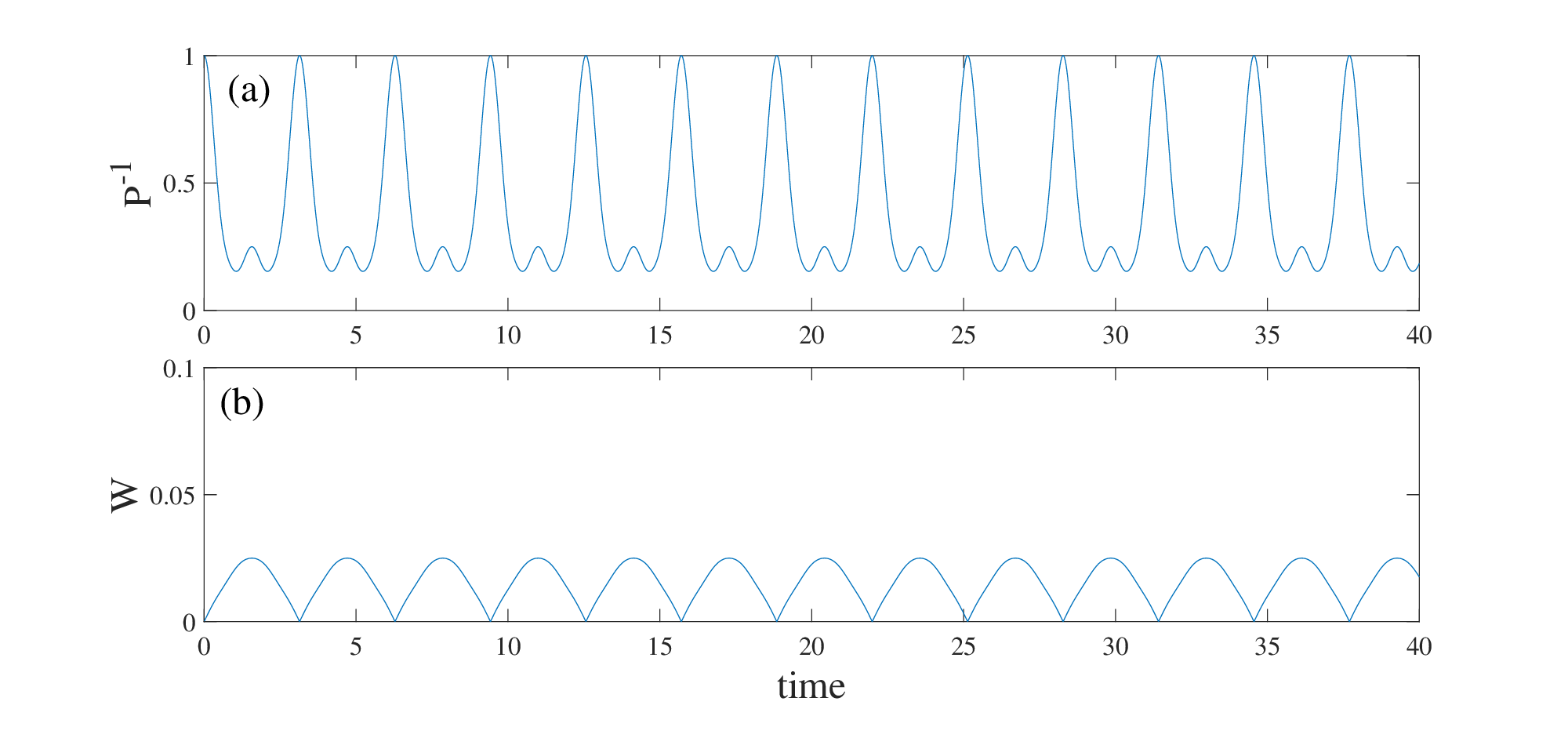}}}
\end{center}
\caption{(a) Inverse participation ratio $P^{-1}$ and (b) average width $W$, as time evolution. The parameters are the same as in Fig. ~\ref{Fig1c}.}\label{FigIPaW1}
\end{figure}

We now consider the dynamics of an injected wave-packet on the central site of a rhombic chain. This wave-packet could be excited with an external laser, and is here modelled with the initial condition $C_{0}=1$ and all other sites initially zero. Fig.~\ref{Fig1c} shows that the initial wave-packet is able to spread to neighboring sites B and C, however, remains caged in the central plaquette (note that the same parameters as in Fig. ~\ref{Fig1a} (b) were used). For further demonstration we show the inverse participation ratio $P^{-1}$ and wave-packet width $W$ in Fig.~\ref{FigIPaW1}. $P^{-1}$ oscillates in a finite range, representing oscillation, which means the localization is occurred. Meanwhile, the width $W$ remains small as shown in Fig.~\ref{FigIPaW1} (b), which is also a signature of wave-packet localization.

\begin{figure}[!tbh]
\begin{center}
\rotatebox{0}{\resizebox *{8.8cm}{6.6cm} {\includegraphics
{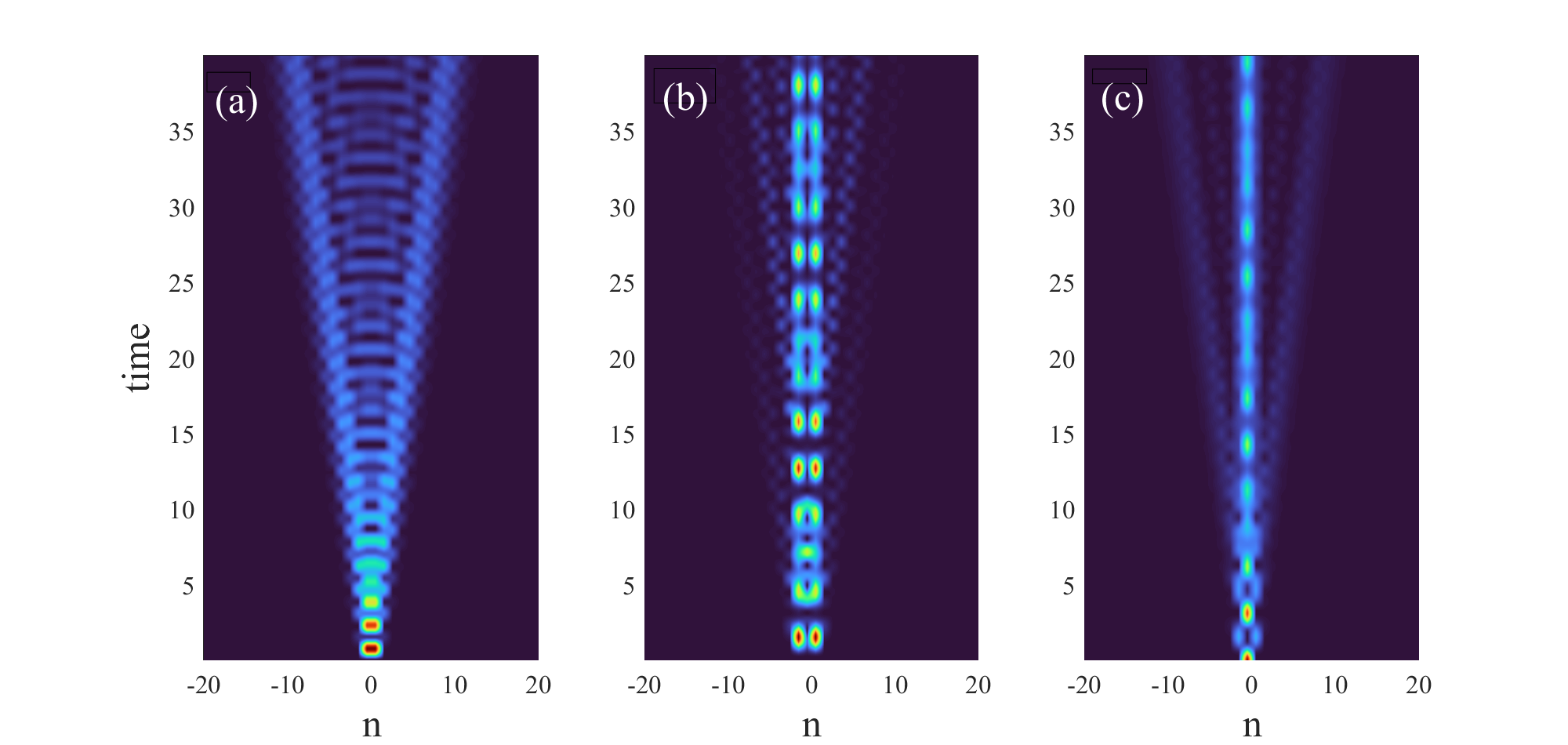}}}
\end{center}
\caption{Dispersive dynamics when the AB caging condition is broken as an example of $\beta_{1}=\pi/3$. With $J=-1$, $\beta_{2}=\pi/4$, $\Delta^{a}_{n}=\Delta^{b}_{n}=\Delta^{c}_{n}=0$, and $\gamma_{a}=\gamma_{c}=\gamma_{b}=0$.}\label{Fig1d}
\end{figure}

\begin{figure}[!tbh]
\begin{center}
\rotatebox{0}{\resizebox *{7.8cm}{5.2cm} {\includegraphics
{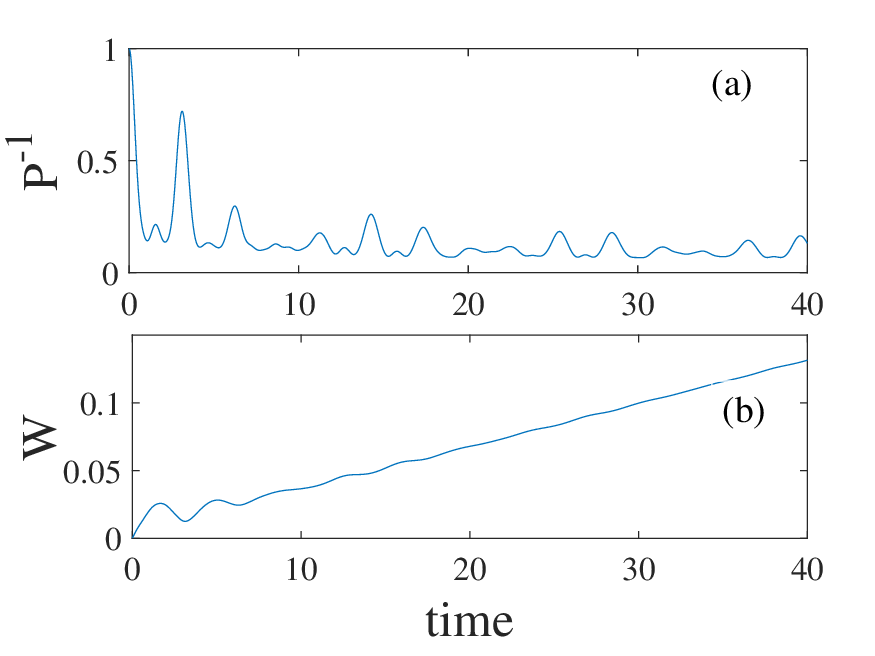}}}
\end{center}
\caption{(a) Inverse participation ratio $P^{-1}$ and (b) average width
$W$, as a function of time $t$. The parameters are the same as in Fig. ~\ref{Fig1d}.}\label{FigIPaW2}
\end{figure}

Fig.~\ref{Fig1d} shows the results of dispersive dynamics for the case of a broken AB caging condition, where we choose the parameters the same as Fig.~\ref{Fig1a} (c). It is clearly displayed that following the time evolution the wave-packet in site $A$, $B$ and $C$ are all dispersive. The other parameters $P^{-1}$ and $W$ are plotted in Fig.~\ref{FigIPaW2}, for further characterizing the localization or delocalization properties. From Fig.~\ref{FigIPaW2}, we can see that for this case, the inverse participation function $P^{-1}$ approaches zero, and the wave-packet width $W$ increases with the time.

\section{AB caging dynamics in the non-Hermitian lattices cases}\label{sec4}

\begin{figure}[!tbh]
\begin{center}
\rotatebox{0}{\resizebox *{8.8cm}{6.8cm} {\includegraphics
{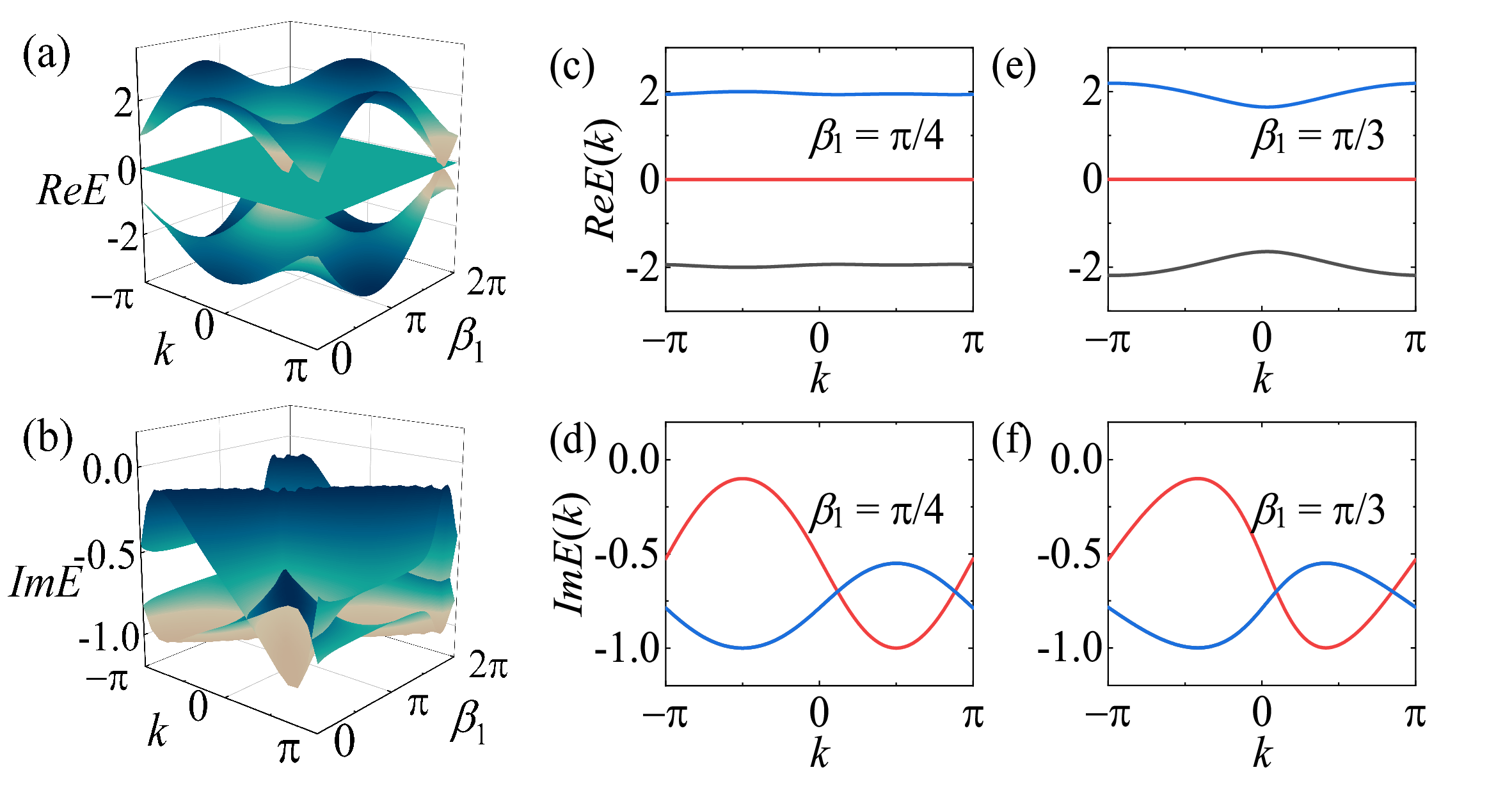}}}
\end{center}
\caption{(a) The real part of energy band of $\mathcal{H}(k)$ for the non-Hermitian case, (c) the real part of energy as a function of $k$ for $\beta_{1}=\pi/4$ and (e) for $\beta_{1}=\pi/3$. (b) The imaginary part of energy, (d) the imaginary part of energy as a function of $k$ for $\beta_{1}=\pi/4$ and (f) for $\beta_{1}=\pi/3$. With $J=-1$, $\beta_{2}=\pi/4$, $\Delta^{a}_{n}=\Delta^{b}_{n}=\Delta^{c}_{n}=0$, $\gamma_{a}=\gamma_{c}=1$ and $\gamma_{b}=0.1$.}\label{Fig2a}
\end{figure}

\begin{figure}[!tbh]
\begin{center}
\rotatebox{0}{\resizebox *{8.0cm}{4.0cm} {\includegraphics
{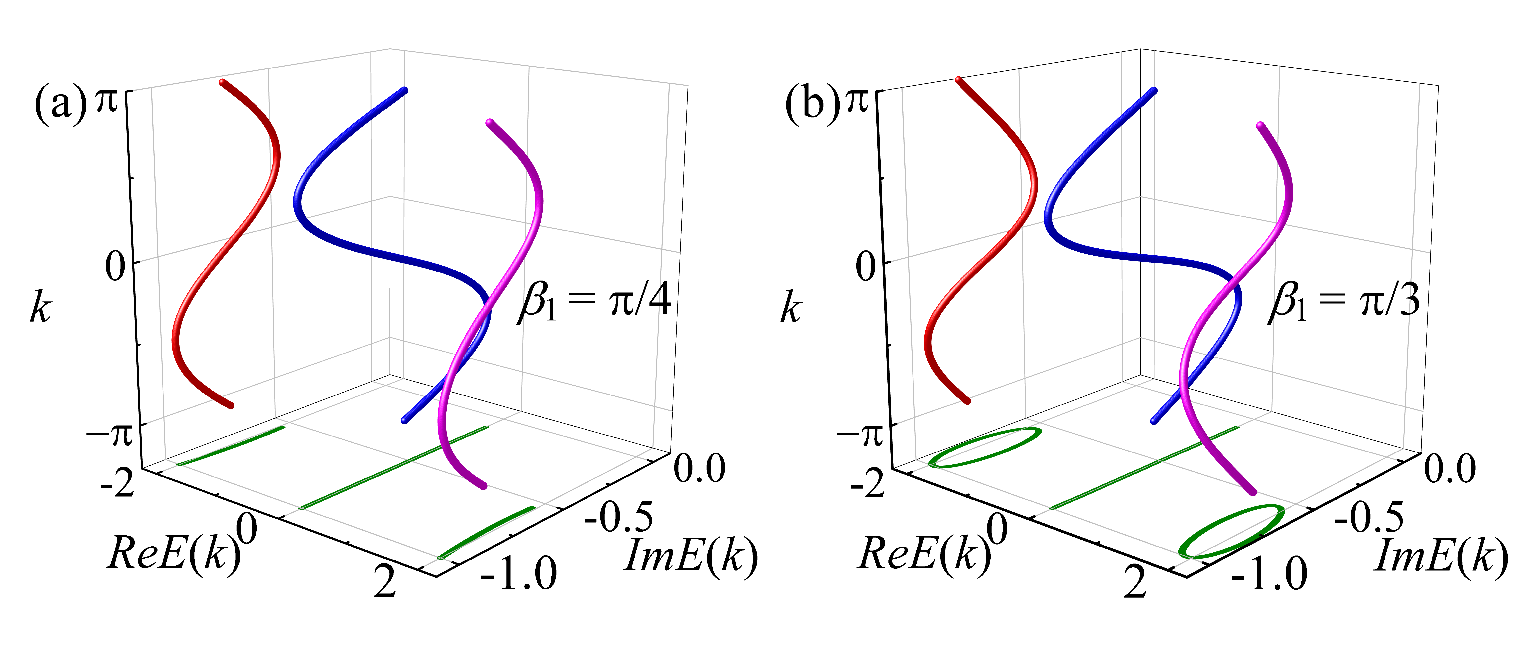}}}
\end{center}
\caption{Complex energies in the ($ReE, ImE, k$) space by connecting the energies at
$k=-\pi$ and $k =\pi$ with (a) $\beta_{1}=\pi/4$ and (b) $\beta_{1}=\pi/3$, respectively. Other parameters are the same as in Fig. ~\ref{Fig2a}.}\label{Fig2b}
\end{figure}

\begin{figure}[!tbh]
\begin{center}
\rotatebox{0}{\resizebox *{8.8cm}{6.6cm} {\includegraphics
{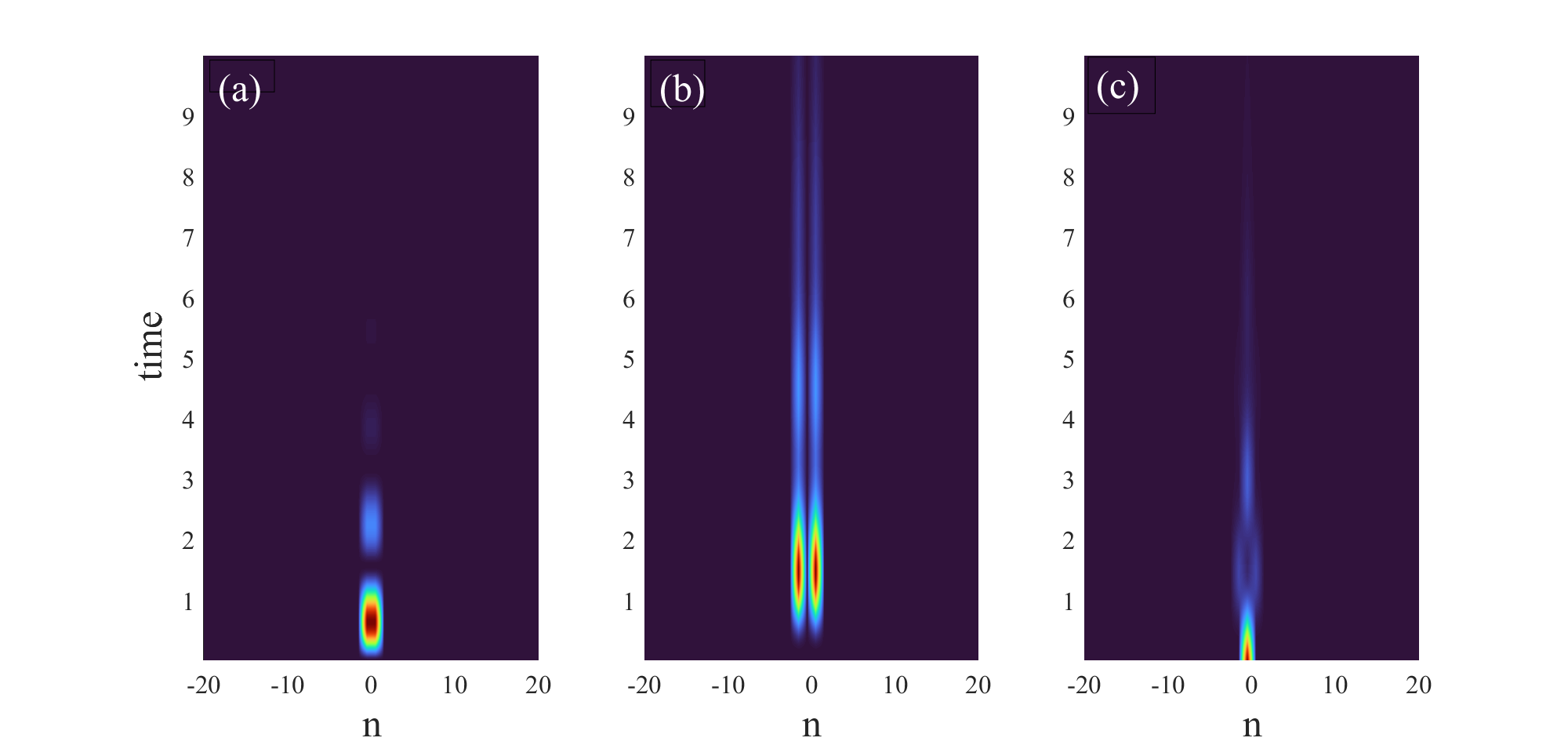}}}
\end{center}
\caption{The decay dynamics for the non-Hermitian cases, the parameter as: $J=-1$, $\beta_{1}$=$\beta_{2}=\pi/4$, $\Delta^{a}_{n}=\Delta^{b}_{n}=\Delta^{c}_{n}=0$, $\gamma_{a}=\gamma_{c}=1$, and $\gamma_{b}=0.1$. }\label{Fig2c}
\end{figure}

In this section, we will discuss the effects of dissipation on the caging dynamics. Indeed, due to the intrinsic non-Hermitian properties of exciton-polariton systems, the dissipation in this system is unavoidable. Further, it is in principle possible that different lattice sites may have different dissipation rates, either through their engineering or partial compensation of dissipation with the application of a non-resonant laser \cite{EWertz}. In this non-Hermitian system the energy will become complex, as shown in Fig. ~\ref{Fig2a}. The results display the real part energy $ReE$ in Fig. ~\ref{Fig2a} (c) and (e), which has a similar dependence on momentum $k$ to the Hermitian case as shown in Fig. ~\ref{Fig1a} (b) and (c). The imaginary part of the energy is plotted in Fig. ~\ref{Fig2a} (d) with $\beta_{1}=\pi/4$ and (f) with $\beta_{1}=\pi/3$, we can see that the imaginary part is little changed by the effective angle $\beta_{1}$. For further study the complex energy band structure, Fig. ~\ref{Fig2b} displays how the real part of energy $ReE$ and imaginary part of energy $ImE$ depends on the momentum $k$. The energy projection on the bottom plane of $(ReE, ImE)$ with green lines shows that they have different topological structure. When the AB caging condition is satisfied as shown in the Fig. ~\ref{Fig2b} (a), the energy bands in the $(ReE, ImE)$ plane form three straight lines. But when the AB caging condition is broken, the energy band structure on the $(ReE, ImE)$ plane will form a two loop structure, corresponding to a non-Hermitian skin effect. Such an effect has been discussed previously in exciton-polariton lattices, in different configurations \cite{SMandal1,SMandal2,XXu,PKokhanchik}. Fig. ~\ref{Fig2c} shows the time evolution of wave-packets, accounting for the presence of dissipation. In contrast to Fig.~\ref{Fig1c}, the oscillations are quickly damped out, however, the main effect of localization remains clearly visible even in this more experimentally realistic configuration.

\section{Inverse Anderson localization phenomenons in the disorder lattices}\label{sec5}
\begin{figure}[!tbh]
\begin{center}
\rotatebox{0}{\resizebox *{8.8cm}{6.6cm} {\includegraphics
{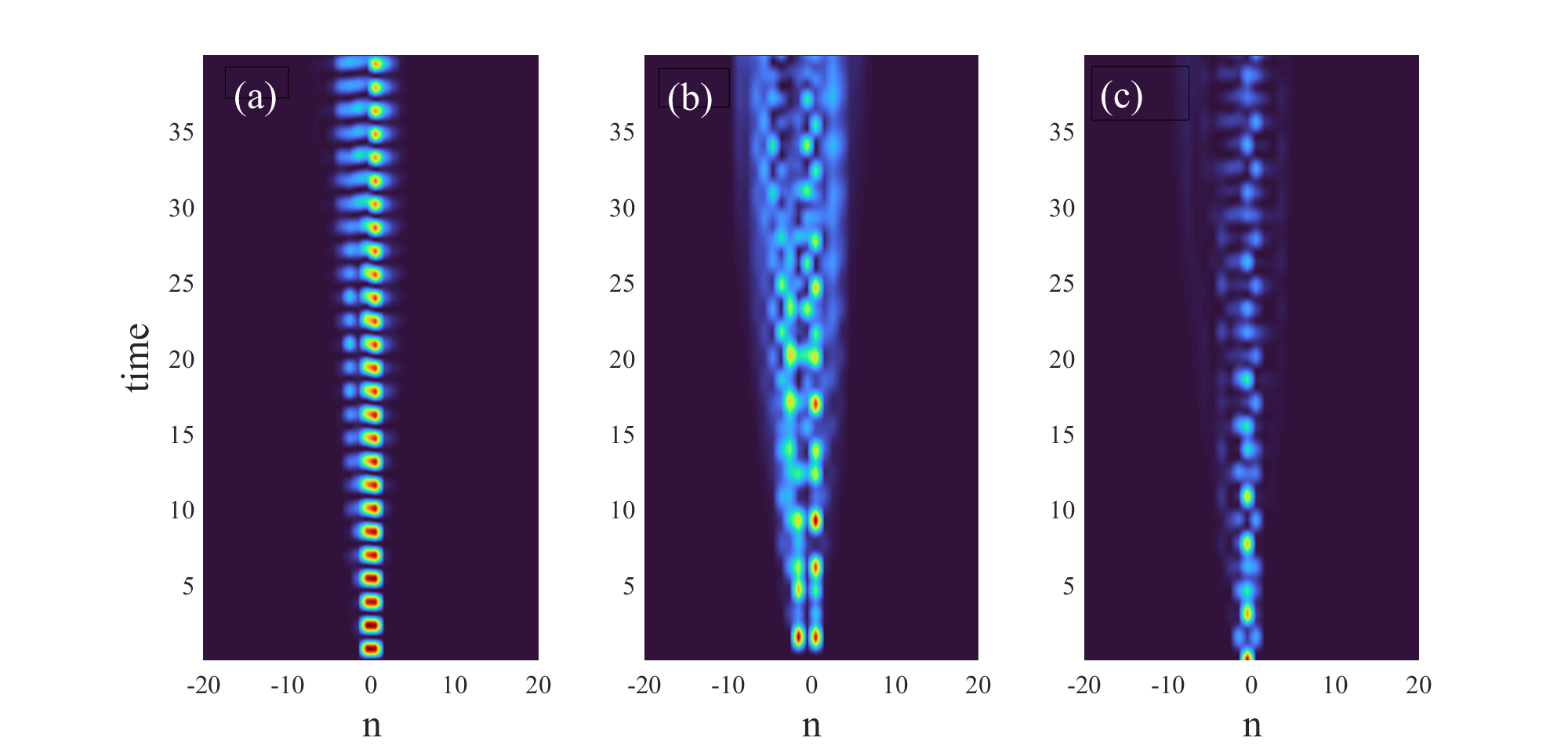}}}
\end{center}
\caption{Time evolution of exciton-polariton condensate when adding the random disorder. Other parameters: $J=-1$, $\beta_{1}$=$\beta_{2}=\pi/4$, $\gamma_{a}=\gamma_{c}=\gamma_{b}=0$.}\label{Fig3a}
\end{figure}

\begin{figure}[!tbh]
\begin{center}
\rotatebox{0}{\resizebox *{6.8cm}{5.2cm} {\includegraphics
{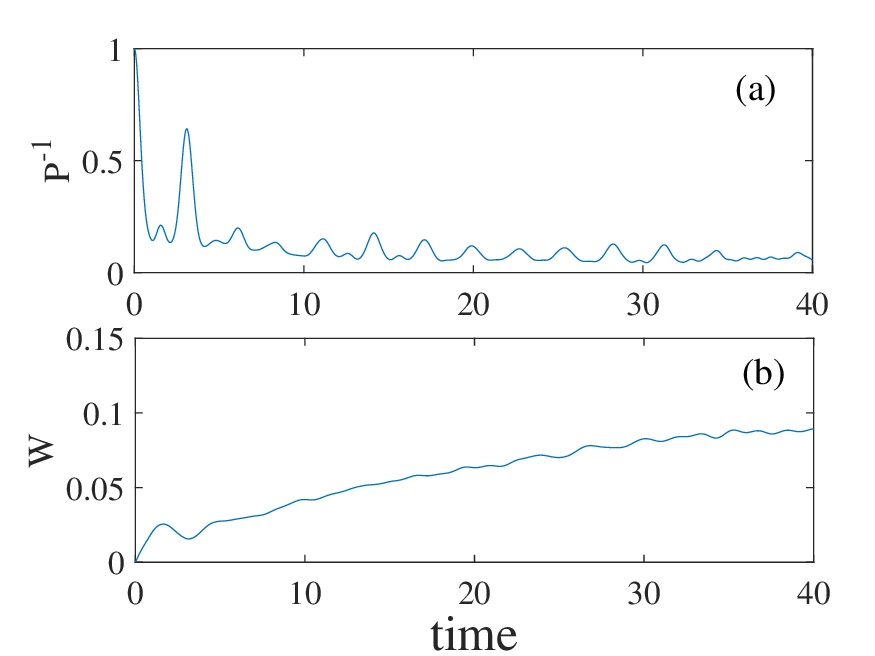}}}
\end{center}
\caption{(a) Inverse participation ratio $P^{-1}$ and (b) average width
$W$, as a function of $t$. The parameters are the same as in Fig. ~\ref{Fig3a}.}\label{FigIPaW3}
\end{figure}

In this part, we will consider the effects of disorder on caging dynamics in the polariton lattice system, i.e, $\Delta^{a}_{n}, \Delta^{b}_{n}, \Delta^{c}_{n}$ not equal to zero but randomly distributed. In such a case, Fig.~\ref{Fig3a} shows that the presence of disorder can lead to wave-packet delocalization. In other words, there is an inverse Anderson localization \cite{Longhi} where disorder actually favours propagation by breaking the AB caging, which is contrast to the typical case of simple lattices where disorder tends to inhibit propagation. Such an effect is further confirmed by the inverse participation parameter $P^{-1}$ and width of wave-packet $W$ as shown in Fig. ~\ref{FigIPaW3} (a) and (b), respectively.

\section{\protect\bigskip Conclusion}\label{conclusion}

In this work, we have theoretically studied the AB caging dynamics in exciton-polariton lattices where artificial gauge fields are induced by spin-orbit coupling. Interestingly, we find that though tuning the flux which is achieved by RDSOC in each plaque, the essential AB caging effects will be achieved in the polariton lattice system. Such effects persist in the presence of polariton losses, which make the system non-Hermitian with a complex band-structure. Moreover, disorder can break the caging dynamics leading to an inverse Anderson transition will also appear in this polariton lattices system. The results open the door for using external gauge fields to manipulate trapping or transportation in exciton-polariton lattices.

\begin{acknowledgments}

Wei Qi was supported by the China Scholarship Council, the National Natural Science Foundation of China under Grant No. 11805116, and the Natural Science Basic Research Plan in Shaanxi Province, China (Grant No. 2023-JC-YB-037). 

\end{acknowledgments}

\end{document}